\documentclass[aps,prl,twocolumn,showpacs,groupedaddress]{revtex4}

\usepackage{amsmath,graphicx,epsfig}

\begin{document}

\title{Shot Noise Detection on a Carbon Nanotube Quantum Dot}

\author{E. Onac}
\author{F. Balestro} \altaffiliation{Present address: Laboratoire Louis
Néel, associ\'e au CNRS, BP 166, F-38042 Grenoble Cedex 9, France}
\author{B. Trauzettel} \altaffiliation{Instituut-Lorentz,
Universiteit Leiden, P.O. Box 9506, 2300 RA Leiden, The Netherlands}
\author{C. F. J. Lodewijk}
\author{L. P. Kouwenhoven} \affiliation{Kavli Institute of
Nanoscience, Delft University of Technology, P.O. Box 5046, 2600 GA
Delft, The Netherlands}

\begin{abstract}
An on-chip detection scheme for high frequency signals is used to
detect noise generated by a quantum dot formed in a single wall
carbon nanotube. The noise detection is based on photon assisted
tunneling in a superconductor-insulator-superconductor junction.
Measurements of shot noise over a full Coulomb diamond are reported
with excited states and inelastic cotunneling clearly resolved.
Super-Poissonian noise is detected in the case of inelastic
cotunneling.
\end{abstract}


\maketitle

The study of shot noise, i.e. non-equilibrium current fluctuations
due to the discreteness of charge carriers, is an important tool for
studying correlations induced in mesoscopic transport by different
types of interactions \cite{buttiker,nazarov}. Current is
characterized by Poissonian shot noise when transport is determined
by an uncorrelated stochastic process. Electron-electron
interactions, such as Coulomb repulsion or resulting from the Pauli
exclusion principle, can correlate electron motion and suppress shot
noise. The noise power density is defined as the Fourier transform
of the current-current correlator
$S_{I}(\omega)=\int_{-\infty}^{+\infty} d\tau \, e^{i\omega \tau} \,
\langle \delta I(t+\tau) \, \delta I(t) \rangle$. This definition is
valid both for positive and negative frequencies $\omega$,
corresponding to energy absorbtion or emission by the device
\cite{gavish,aguado,schoelkopf}. When $|eV| \gg |\hbar \omega|,
k_{B}T$ ($V$ is the voltage bias and $T$ the temperature), shot
noise dominates over other types of noise and the power density has
a white spectrum that can be expressed as
$S_I(-\omega)=S_I(\omega)=FeI$. Here, $I$ is the average current and
the Fano factor, $F$, indicates the deviation from Poissonian shot
noise for which $F=1$. If the noise detector can not distinguish
between emission and absorption processes, a symmetrized version
$S_{I}^{symm}(\omega)=S_{I}(\omega)+S_{I}(-\omega)$ is used. The
Schottky formula $S_{I}^{symm}=2eI$ refers to this symmetrized case.

For electron transport through a quantum dot (QD) shot noise can be
either suppressed or enhanced with respect to the Poissonian value.
First, for resonant tunneling, when a QD ground state is aligned
between the Fermi levels in the leads, the Fano factor can vary
between 1/2 and 1. The exact value is determined by the ratio of
tunneling rates between the dot and the two leads \cite{theoryQD}.
For strongly asymmetric barriers transport is dominated by the most
opaque one and shot noise is Poissonian. If the barriers are
symmetric, the resonant charge state is occupied 50\% of the time
and a $F=1/2$ shot noise suppression is predicted. Second, when the
QD is in Coulomb blockade, first-order sequential tunneling is
energetically forbidden. Transport can still occur via cotunneling
processes \cite{silvano}, elastic or inelastic. These are second
order processes, with a virtual intermediate state, allowing
electron transfer between the leads. The elastic process leaves the
QD in its ground state and transport is Poissonian. Inelastic
cotunneling switches the system from a ground to an excited state
and can lead to super-Poissonian noise with a Fano factor up to
$F=3$ \cite{co-tunnoise}. Experiments have shown shot noise
suppression due to Coulomb blockade \cite{schoenenberger,haug}, but
no experimental results exist on shot noise enhancement in the
inelastic cotunneling regime. Here, we present the detection of
noise, generated by a carbon nanotube quantum dot (CNT-QD). Excited
states and inelastic cotunneling are clearly resolved in the noise
measurements. For inelastic cotunneling we find super-Poissonian
shot noise.


\begin{figure}
\includegraphics[width=8.3cm]{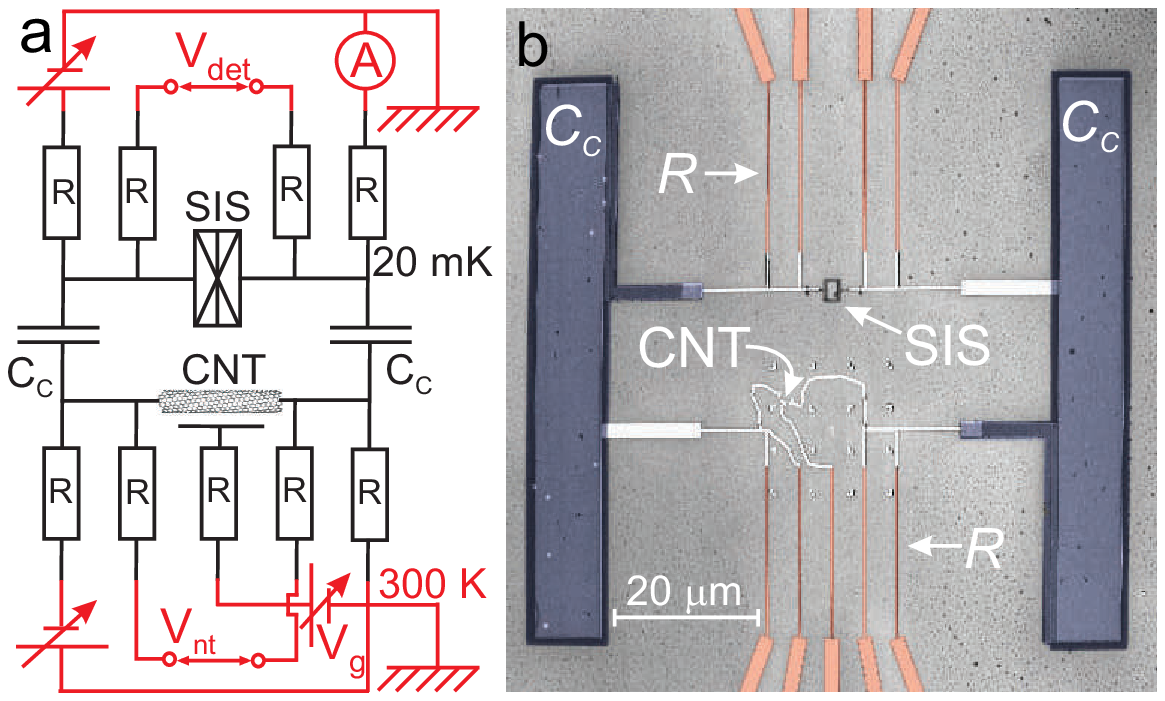} \\
\includegraphics[width=4.3cm]{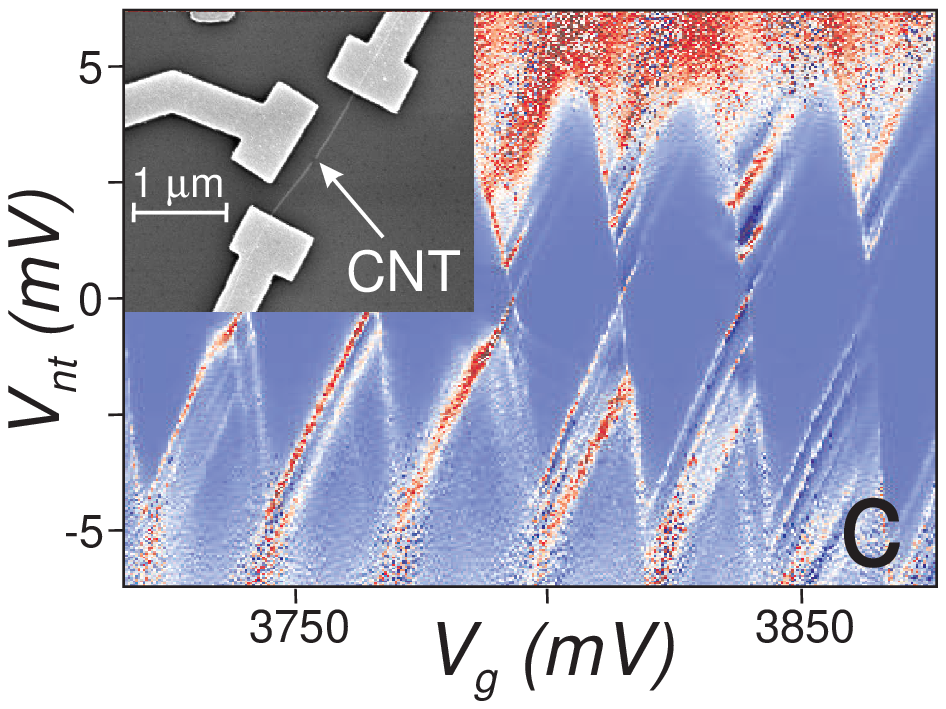}
\includegraphics[width=3.78cm]{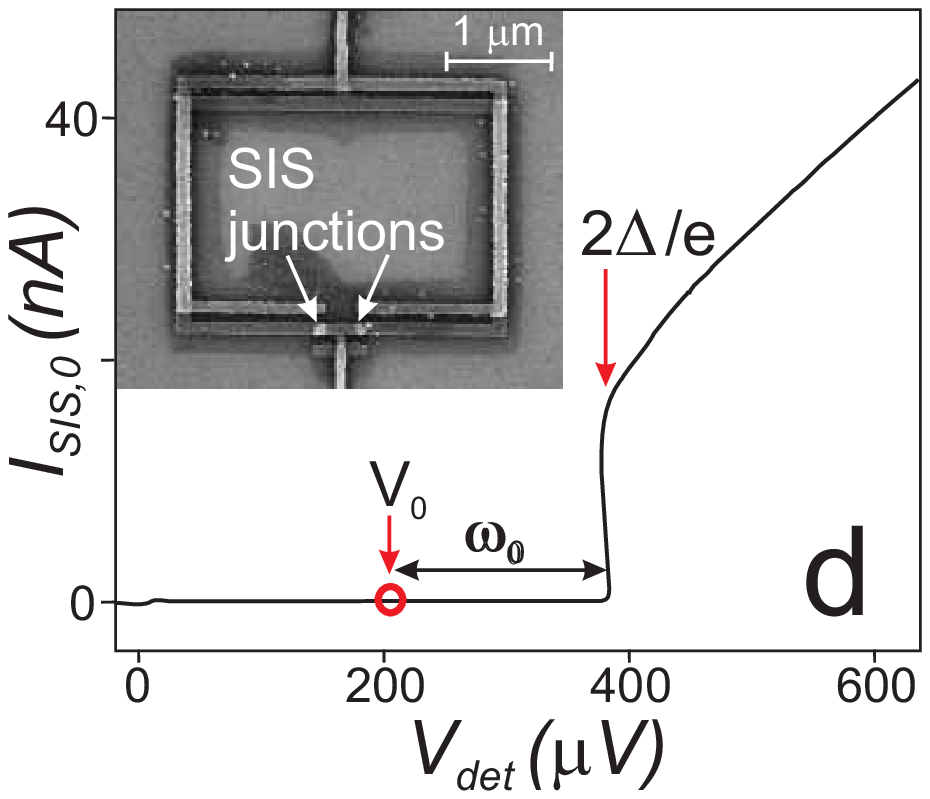}
        \caption{Schematic drawing of the circuit \textbf{(a)} and SEM picture of the sample \textbf{(b)}. The SIS detector is coupled, using two on-chip capacitances $C_C$
(blue part), to the CNT-QD (the curly white lines are the custom
made contacts) and is cooled to 20 mK. Four contact lines (red) are
used for both the CNT and the detector. An additional contact on the
CNT side is used as a side gate (see inset in (c)). All lines
incorporate an on-chip impedance $R$ to prevent the high frequency
signal from leaking via parasitic capacitances in the leads. These
are thin Pt wires (25x0.1x0.02 $\mu m$) with a resistance between 2
and 2.25 $k\Omega$, measured at 4K. \textbf{(c)} Carbon nanotube
$dI_{nt}/dV_{nt}$ density plot shows standard Coulomb diamonds.
\textit{Inset}: SEM picture of the CNT with contacts and a side
gate. \textbf{(d)} $I-V$ characteristic of the detector SIS junction
in absence of noise. Current is zero in the superconducting gap
region and is determined by the normal state resistance, $R_N=11.3$
k$\Omega$, for $V_{det}>2\Delta/e$. \textit{Inset}: SEM picture of
the detector with two SIS junctions. This SQUID geometry allows us
to suppress the supercurrent by means of an external magnetic
field.}
        \label{fig:SISscheme}
\end{figure}

We use an on-chip noise detector consisting of a
superconductor-insulator-superconductor (SIS) junction. Noise
generated in the CNT-QD leads to photon assisted tunneling between
the superconducting electrodes of the SIS detector (see
Fig.~\ref{fig:SISscheme}(a),(b)). This causes a change in the
detector current, that contains information about the spectral power
of noise \cite{deblock}. The frequency range of the SIS detector is
determined by the superconducting gap $\Delta$ (5-90 GHz for Al).

Sample fabrication necessitates five steps of \textit{e}-beam
lithography and material deposition for the different circuitry
parts \cite{thesis}. In intermediate steps CVD deposition
\cite{NTgrowth} and AFM imaging are used for growing and locating
the nanotubes. A 20 nm Pt layer is deposited for contacts and the
lower plate of the coupling capacitances (see
Fig.~\ref{fig:SISscheme}(b)). For the insulating layer of the
capacitances we use 40 nm of SiO. In a last step, angle evaporation
with an intermediary oxidation step is employed to deposit the Al
tunnel junctions for the SIS detector and the upper plate of the
capacitances.

Current fluctuations in the CNT-QD, $S_{I}(\omega)$, induce, via the
coupling capacitances, voltage fluctuations across the detector,
$S_{V}(\omega)$. The detector current in absence of noise,
$I_{SIS,0}(V_{det})$, has a step-like shape (see
Fig.~\ref{fig:SISscheme}(d)), which is modified by $S_V(\omega)$
into $I_{SIS}(V_{det})$. More specifically, the emission side of the
spectrum $S_{V}(-\omega)$ induces a change
$I_{det}=I_{SIS}-I_{SIS,0}$ in the superconducting gap region
($0<V_{det}<2\Delta/e$) \cite{thesis}:
\begin{equation}
        I_{det}(V_{det})= \int_{0}^{+\infty} \negthickspace \negthickspace d\omega \, \left( \frac{e}{\hbar\omega}\right)^2 \, \frac{S_{V}(-\omega)}{2\pi} \; I_{SIS,0} \negthickspace \left( V_{det}+\frac{\hbar \omega}{e} \right)
        \label{eq:Ipatsimple}
\end{equation}
Note that $I_{SIS,0}(V_{det})\neq 0$ only for $V_{det}>2\Delta/e$.
If we consider a detector voltage $V_{det}=V_0$ (see example in
Fig.~\ref{fig:SISscheme}{d}) then only frequencies above
$\omega_{0}=(2\Delta/e-V_{0})/\hbar$ contribute to the detector
current. This means that each point on the detector curve
$I_{det}(V_0)$ represents noise detection over a bandwidth
$(\omega_{0},\infty)$. However, contributions from different
frequencies are normalized as $S_{V}(\omega)/\omega^2$, leading to
smaller changes in the detector current for higher frequencies.
Finally, $S_V$ is related to the CNT-QD current fluctuations by
$S_{V}(-\omega)=S_{I}(-\omega)|Z(\omega)|^2$, with the
transimpedance $Z(\omega)$ being determined by the coupling
circuitry.

In the regime $|eV_{nt}| \gg |\hbar \omega|, k_{B}T$ ($V_{nt}$ is
the CNT bias voltage) shot noise dominates over other types of
noise. Here, the power density is proportional to the average
current, $S_{I}\sim I$, and frequency independent
$S_{I}(\omega)=S_{I}(-\omega)=S_{I}(0)=constant$ (i.e. white
spectrum). Eq.(\ref{eq:Ipatsimple}) can then be written as:
\begin{equation}
        I_{det}(V_{det})= \frac{S_{I}(0)}{e}\; \kappa \! \left( Z, I_{SIS,0}, V_{det} \right)
        \label{eq:Ipatsimple2}
\end{equation}
with $\kappa \!= \int_{0}^{+\infty} d\omega \, \frac{e^3}{2\pi
(\hbar\omega)^2} \, |Z(\omega)|^{2} \, I_{SIS,0} \! \left(
V_{det}+\frac{\hbar \omega}{e} \right)$ a function that depends on
transimpedance, detector $I-V$ characteristic in the absence of
noise, and detector bias. Eq.(\ref{eq:Ipatsimple2}) is valid in
general, for any white noise source that is coupled to the SIS
detector junction.


The single wall carbon nanotube (CNT) has a length of 1.6 $\mu$m
between the contacts and a side gate is used to change its
electrical potential (see Inset of Fig.~\ref{fig:SISscheme}(c)).
From the observation that we can induce both electron and hole
transport at room temperature, we conclude that we have a small gap
semiconductor CNT. After cooling to 20 mK, the conductance,
$dI_{nt}/dV_{nt}$, density plot, as a function of applied bias,
$V_{nt}$, and gate voltage, $V_{g}$, shows closing Coulomb diamonds,
implying that one quantum dot is formed (see
Fig.~\ref{fig:SISscheme}(c)). Excited states are clearly stronger
for one direction (parallel to one side of the Coulomb diamonds)
indicating asymmetric tunnel barriers to the leads. From the size of
the larger Coulomb diamonds, we estimate the addition energy $\delta
+ E_{C}\approx 4$ mV, with $\delta \approx 1$ meV the orbital energy
and $E_{C} \approx 3$ meV the charging energy. The value for
$\delta$ is consistent with the figure expected for a quantum dot
formed by barriers at the contacts.

\begin{figure}[t]
                \includegraphics[width=4.2cm]{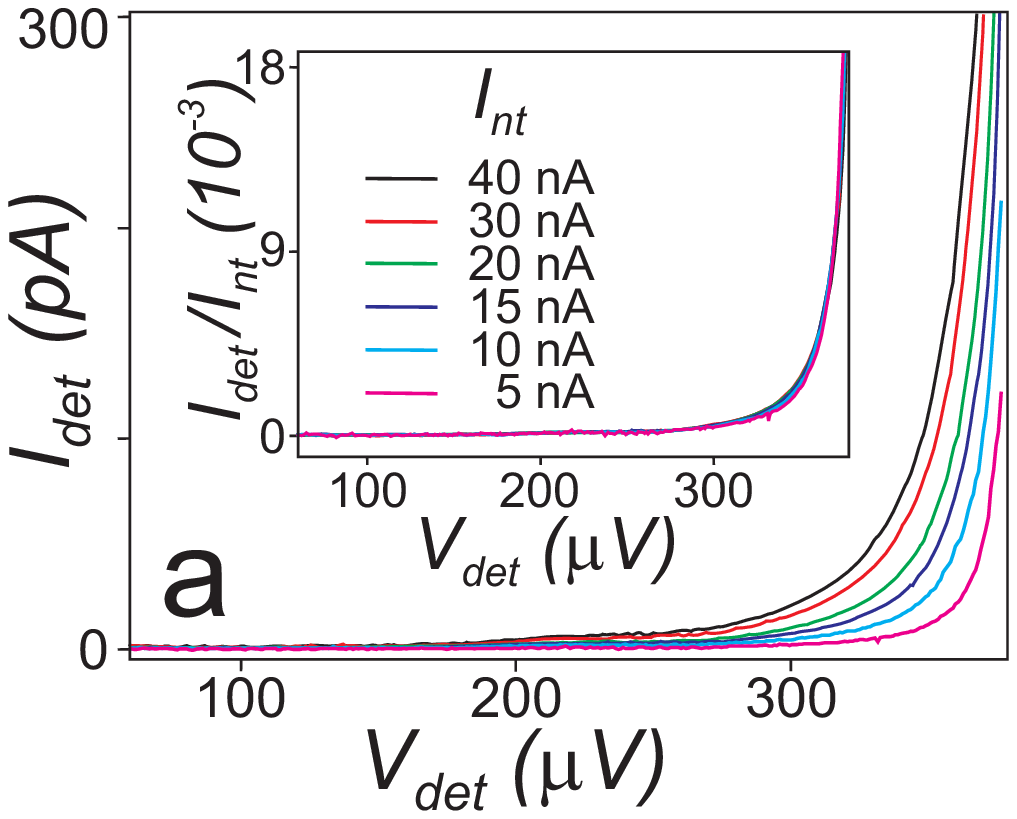}
                \includegraphics[width=4.2cm]{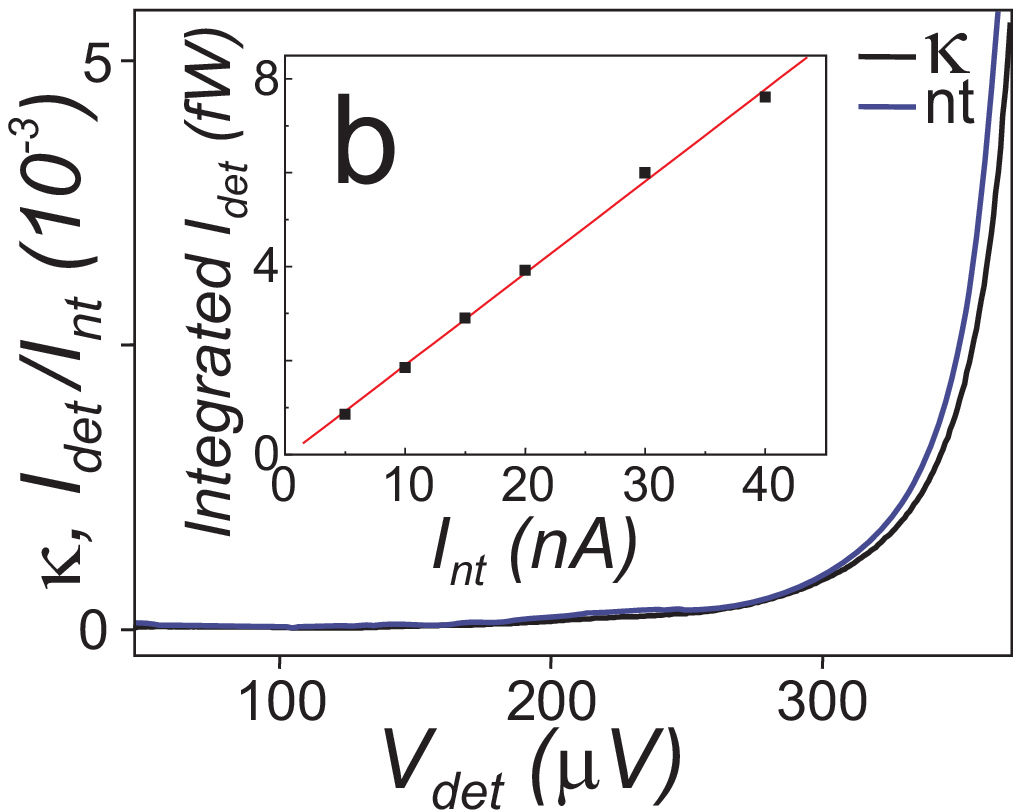}
        \caption{\textbf{(a)} Detector signal, $I_{det}$, as a function of detector bias, $V_{det}$,
 for several CNT current bias values, $I_{nt}$. The maximum current
 bias $I_{nt}=40$ nA corresponds to a voltage $V_{nt}=11$ mV.
 \textit{Inset}: same detector curves normalized to CNT current.
\textbf{(b)} Averages of normalized detector curves for the
calibrating SIS-SIS, respectively SIS-CNT sample. \textit{Inset}:
Integrated detector signal, $\int I_{det} \, dV_{det}$ (for 140
$\mu$V $ \leq V_{det} \leq $ 370 $\mu$V), versus CNT current,
$I_{nt}$.}
        \label{fig:CBNTcuts&nc}
\end{figure}

We fix the gate voltage at a Coulomb peak and current bias the
nanotube. The detector signal, $I_{det}$, presented as a function of
the detector bias in Fig.~\ref{fig:CBNTcuts&nc}, is increasing with
the CNT current, $I_{nt}$. The fact that the normalized curves
$I_{det}/I_{nt}$ are all identical, over this range of $I_{nt}$ (see
inset of Fig.~\ref{fig:CBNTcuts&nc}(a)), proves that we are indeed
measuring white shot noise. This is also apparent from the inset of
Fig.~\ref{fig:CBNTcuts&nc}(b), showing that the integrated detector
signal depends linearly on the nanotube current.

Since the power spectral density can be expressed as
$S_{I}=FeI_{nt}$, the normalized curve can be written as
$I_{det}/I_{nt} = F \, \kappa \left( Z, I_{SIS,0}, V_{det} \right)$.
We determine the circuit calibration function $\kappa$ by using a
separate sample, in which well-known Poissonian noise is generated
\cite{deblock}. This calibration sample is fabricated simultaneously
with the CNT sample, but with an SIS junction as a noise source. The
obtained calibration curve is presented in
Fig.~\ref{fig:CBNTcuts&nc}(b). We also plot there the normalized
curve $I_{det}/I_{nt}$, averaged for CNT currents between 5 nA and
40 nA. The two curves have similar amplitudes, meaning that, for a
given value of the current through the source, the detector signal
is the same for the two samples. This indicates a Fano factor close
to the Poissonian value $F=1$ \cite{calibration1} for the high bias
regime of the CNT. Based on these considerations, we use below the
normalized curve, $\kappa_{nt}=I_{det}/I_{nt}$, in
Fig.~\ref{fig:CBNTcuts&nc}(b) as a calibration curve. We estimate
that the deviation of the Fano factor from the $F=1$ value is less
than 12\% \cite{thesis}. Our calibration allows for detection of
changes in the Fano factor within this error bar.

\begin{figure}[t]
                \includegraphics[width=8.3cm]{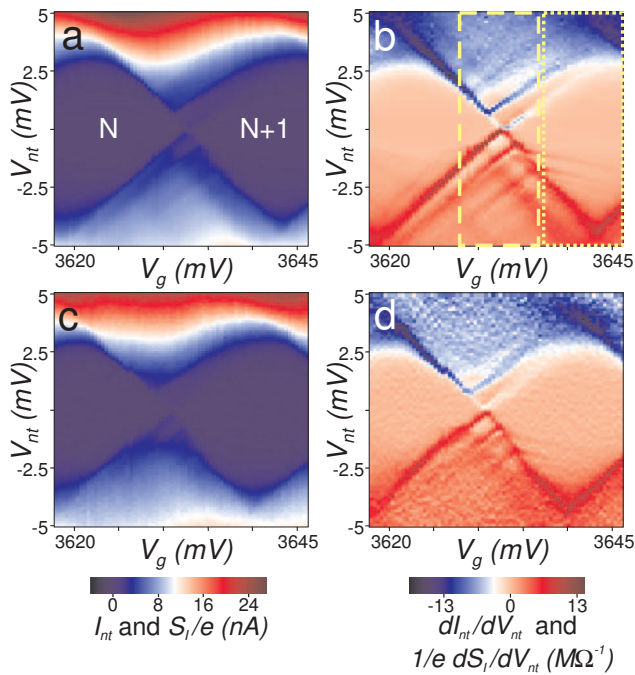}
        \caption{Density plots for DC (\textbf{(a)} and \textbf{(b)}) and
noise measurements (\textbf{(c)} and \textbf{(d)}) as a function of
nanotube bias, $V_{nt}$, and gate voltage, $V_{g}$. Standard DC
measurement of current, $I_{nt}$, and conductance,
$dI_{nt}/dV_{nt}$, are presented in \textbf{(a)} and \textbf{(b)}
for two adjacent Coulomb diamonds. The diamonds correspond to a
fixed number of electrons ($N$, respectively $N+1$) on the quantum
dot. Noise power, $S_I$, obtained according to Eq.(\ref{eq:FI}),
together with its derivative, $dS_I/dV_{nt}$, are presented in
\textbf{(c)} respectively \textbf{(d)}. Both are normalized to the
electron charge such that we can use the same color scale for DC and
noise measurements.}
        \label{fig:CD_ac_noise}
\end{figure}

We now focus on the two adjacent Coulomb diamonds in
Fig.~\ref{fig:CD_ac_noise}(a), with its derivative plotted in
Fig.~\ref{fig:CD_ac_noise}(b). Subsequently, we use the SIS detector
to measure shot noise. We fix the gate voltage, $V_{g}$, and measure
the detector current with finite ($I_{SIS}$) and zero ($I_{SIS,0}$)
nanotube bias voltage, $V_{nt}$. Then the values for $V_{g}$ and
$V_{nt}$ are changed and the noise measurements repeated. In this
way we obtain the detector signal $I_{det}=I_{SIS}-I_{SIS,0}$ over
the entire range of the Coulomb diamond.

We sweep the detector bias $V_{det}$ only over a limited interval
$(V_{det}^i,V_{det}^f)$ of the superconducting gap region, where the
detector is most sensitive \cite{bias_sweep}. We obtain the noise
power over this interval from
\begin{equation}
        S_I/e={{\displaystyle\int_{V_{det}^i}^{V_{det}^f} {I_{det}(V_{det}) \; dV_{det}}}} \left/ {{\displaystyle\int_{V_{det}^i}^{V_{det}^f} \kappa_{nt} \; dV_{det}}
        }\right.
        \label{eq:FI}
\end{equation}
The resulting density plots for noise (see
Fig.~\ref{fig:CD_ac_noise}(c) and (d)) are in good correspondence
with the ones from the standard DC measurement
(Fig.~\ref{fig:CD_ac_noise}(a) and (b)). This is expected, as
changes in $I_{nt}$ also give changes in $S_I \propto I_{nt}$. There
is a small shift in gate voltage values between the noise and the DC
measurement (due to the long measurement time for noise detection).
Excited states, as well as inelastic cotunneling signal inside the
Coulomb diamond, are clearly resolved for both types of
measurements.

\begin{figure}[t]
\includegraphics[width=4.1cm]{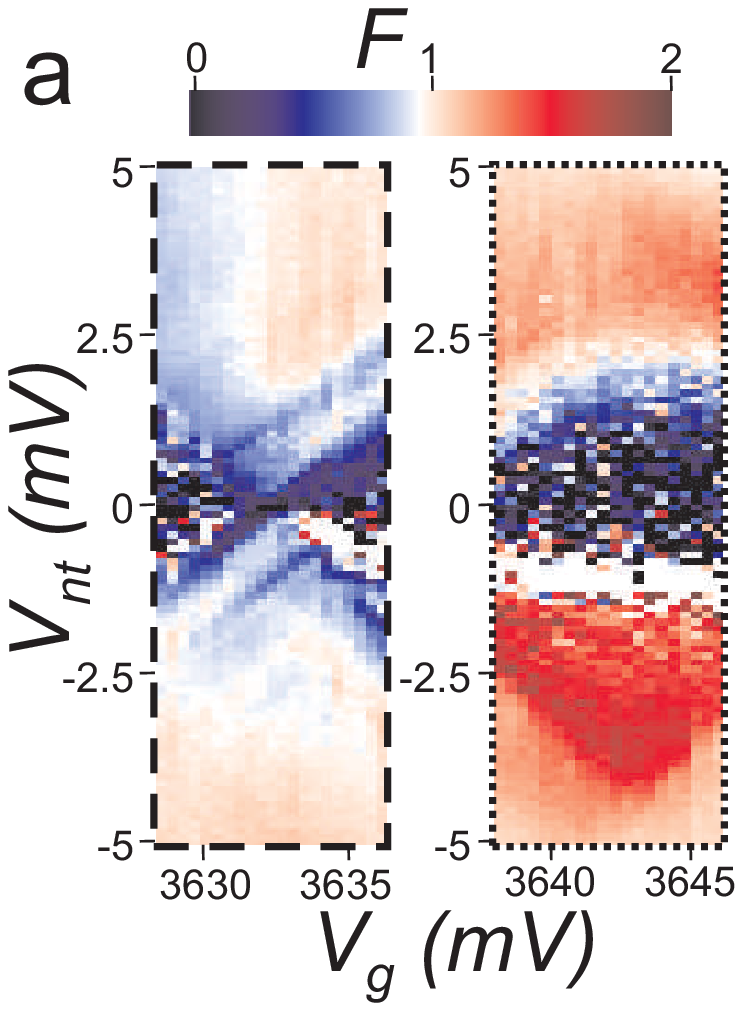}
\includegraphics[width=4.3cm]{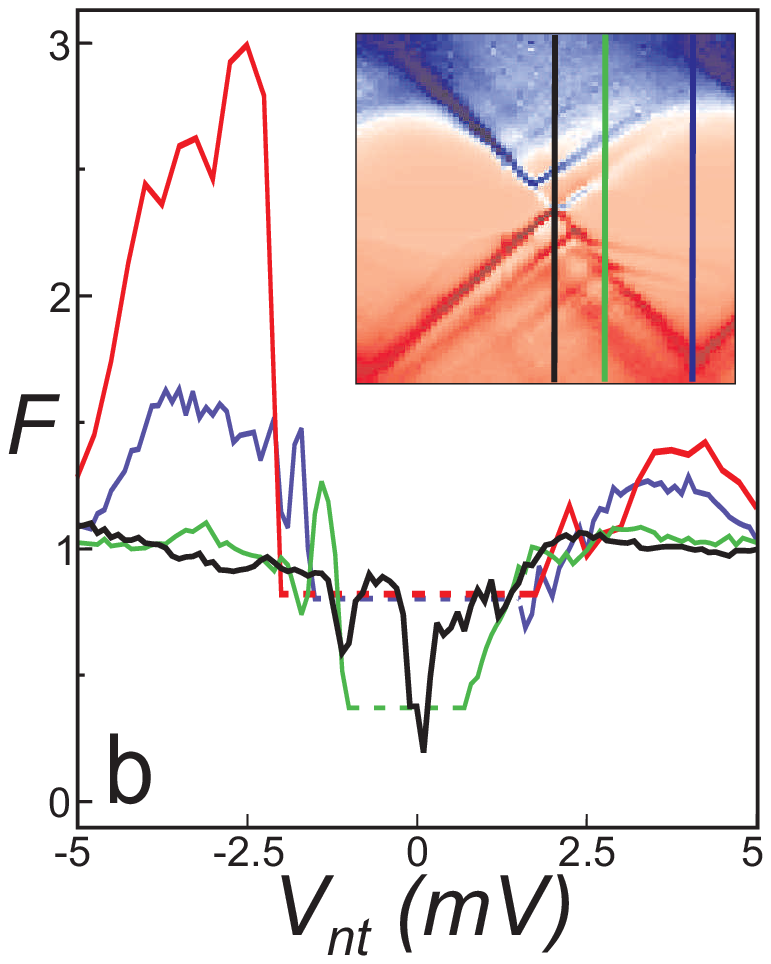}
        \caption{\textbf{(a)} Fano factor density plots corresponding to the two parts of the
Coulomb diamond indicated in Fig.~\ref{fig:CD_ac_noise}(b).
\textbf{(b)} Individual Fano factor curves determined for gate
voltages indicated in the inset density plot. The red curve was
measured in a different Coulomb diamond. $F>1$ indicates
super-Poissonian noise corresponding to inelastic cotunneling. For
CNT currents $I_{nt}<$ 150 pA (the dashed part of the curves) no
excess noise can be measured, with the sensitivity of our detection
scheme.}
        \label{fig:Fano}
\end{figure}

The density plot for the Fano factor can be obtained by dividing the
plots in Fig.~\ref{fig:CD_ac_noise}(c) and \ref{fig:CD_ac_noise}(a),
after careful alignment to correct for small gate shifts. In this
way we get the Fano factor values for specific regions outlined by
the dashed and dotted lines in Fig.~\ref{fig:CD_ac_noise}(b). These
values, presented in Fig.~\ref{fig:Fano}(a), indicate a noise
suppression ($F<1$) at the closing of the diamond (dashed part) and
an enhancement of noise ($F>1$) in the regime of inelastic
cotunneling (dotted part). Fano factor curves are also individually
determined and plotted in Fig.~\ref{fig:Fano}(b).

We first consider the situation when the QD is outside Coulomb
blockade (left part of Fig.~\ref{fig:Fano}(a) and black curve in
Fig.~\ref{fig:Fano}(b)). For small biases, close to the diamond
crossing, we find that noise is suppressed, $F<1$. This indicates
that transport is not dominated by a single barrier and resonance
effects matter. At large biases (where also $I_{nt}>$5 nA) we detect
Poissonian shot noise, in agreement with the calibration procedure.
Thus, measurements in the sequential tunneling regime are consistent
and prove that our detection scheme is reliable.

We now look at the region inside the Coulomb diamond, where
transport occurs via cotunneling (see right part of
Fig.~\ref{fig:Fano}(a)). First, for elastic cotunneling, no noise is
measured ($F \approx 0$ in the dark-blue region). This is a second
order process, in which an electron is transferred between the
leads, via an intermediate virtual state. The electron has a very
short dwell time and leaves the dot in its ground state. Subsequent
elastic cotunneling events are completely uncorrelated and
Poissonian shot noise is predicted, i.e. $F=1$. However, our signal
is obtained after subtracting the detector $I-V$ in the absence of
device bias: $I_{det}=I_{SIS}-I_{SIS,0}$. We only measure the excess
noise (the noise induced by the CNT bias). In the regime of elastic
cotunneling $I_{nt}$ is too small ($<150$ pA) to give a measurable
contribution to the excess noise, and our substraction procedure
yields $F \approx 0$.

Finally we consider the inelastic cotunneling regime. The green
curve in Fig.~\ref{fig:Fano}(b), taken at a gate value where
inelastic cotunneling sets in, shows a small region in $V_{nt}$ with
super-Poissonian noise. The blue curve indicates an increase of the
region with $F>1$. Measurements of super-Poissonian noise, due to
inelastic cotunneling, were also performed for other Coulomb
diamonds (see red curve in Fig.~\ref{fig:Fano}(b)), showing a very
pronounced Fano factor enhancement. Super-Poissonian noise can occur
when two channels, with different transparencies, are available for
transport \cite{co-tunnoise,safonov}. If only one can be open at a
time, electrons are transferred in bunches whenever transport takes
place through the more transparent channel. Such conditions are met
by a quantum dot in the inelastic cotunneling regime. In the ground
state, current is blocked due to Coulomb interaction. Still, if the
bias is larger than splitting between the ground and the first
excited state, a second order, inelastic tunneling processes can
take place and an electron is transferred from one lead to the
other. The inelastic cotunneling leaves the dot in the excited
state. The electron can subsequently either tunnel out or relax to
the ground state and block again the current. Thus, depending on the
tunneling rate through the excited state and the relaxation rate, we
can distinguish two regimes. If the electron relaxes to the ground
state, we are in the weak cotunneling regime. For noise, this is
equivalent to elastic cotunneling (the electron always relaxes and
tunnels out from the ground state) and leads to Poissonian noise
$F=1$. If relaxation is slow and transport takes place through the
excited state (strong cotunneling regime), electrons are transferred
in bunches and the noise becomes super-Poissonian. For inelastic
cotunneling we measure $F>1$ , showing that we are in the strong
cotunneling regime. Still relaxation processes play an important
role and lead to a Fano factor smaller than the maximum $F=3$
predicted value.


Noise measurement over the entire Coulomb diamond region of a QD is
reported for the first time. Features present in the standard
current measurements, including excited states and inelastic
cotunneling, are clearly resolved in noise measurements. This
confirms the high sensitivity and versatility of our detection
scheme. Super-Poissonian noise ($F>1$), corresponding to inelastic
cotunneling, is detected, also for the first time.

We are grateful to P. Jarillo-Herrero for assistance in the CNT
fabrication. We acknowledge the technical support from R. Schouten
and B. van der Enden. Financial support was provided by the Dutch
Organisation for Fundamental Research (FOM).


\begin{thebibliography}{15}
\bibitem{calibration1}
{A higher curve for the CNT sample is in agreement with a larger CNT
impedance $R_{S}=200$ k$\Omega$ compared with $R_{N}=15$ k$\Omega$
of the SIS junction used for calibration \cite{thesis}.}
\bibitem{bias_sweep}
{The applied detector bias is between $350$ $\mu$V and 400 $\mu$V.
Due to series resistances in the circuit, this results in a real
detector bias between $V_{det}^i\simeq 333$ $\mu$V and
$V_{det}^f\simeq 370$ $\mu$V. The corresponding cut-off frequencies
$\omega_{i}\simeq 13.2$ GHz, respectively $\omega_{f}\simeq 4.3$
GHz, represent the lower limit for the detection bandwidth.}
\end{thebibliography}
\end{document}